%% file: symmetry_v3.tex
\documentclass[11pt]{article}  
 \pdfoutput=1 
\usepackage{jheppub}
\usepackage{titlesec}
\usepackage{tikz, amsmath}  
\usetikzlibrary{calc} 
\usepackage{bbm}  
\usepackage{xcolor} 
\usepackage{dsfont} 
\usepackage{subfigure} 
\usepackage{hyperref}  
\usepackage{microtype} 
\usepackage{caption} 
\usepackage{amsmath}  
\usepackage{float} 
\usepackage{graphics}  
\usepackage{psfrag} 
\usepackage{color} 
\usepackage{slashed}  
\usepackage{subfigure}
\usepackage{graphicx} 
\usepackage{array,multirow} 
\usepackage{amsfonts}
\usepackage{amssymb}
\usepackage{mathtools}
\usepackage{amsmath}
\usepackage{amsfonts}
\usepackage{mathrsfs}
\usepackage{multirow}
\usepackage{array}
\usepackage{textcomp} 
\usepackage{phaistos} 
\usepackage[utf8]{inputenc}
\usepackage{pifont} 
\usepackage{pgf}
\usepackage{enumitem}
\usepackage{mathtools}

\usepackage{subfigure}
\usepackage{multirow}

\newcommand{\beq}{\begin{eqnarray}}
\newcommand{\eeq}{\end{eqnarray}}

\newcommand{\bmp}{\noindent\begin{minipage}{16cm}}
\newcommand{\emp}{\end{minipage}\vskip 7mm}

\usepackage{dcolumn}
\usepackage{bm}
\usepackage{slashed} 

\usepackage{epsfig}

\newcommand{\bea}{\begin{eqnarray}}
\newcommand{\eea}{\end{eqnarray}}

\newcommand{\ba}{\begin{eqnarray}}
\newcommand{\ea}{\end{eqnarray}}

\newcommand{\de}{\partial} 

\title{Diffeomorphism on-shell breaking from radion stabilization}  

\author{Haiying Cai}
\emailAdd{hcai@korea.ac.kr}
\affiliation{ Department of Physics, Korea University, Seoul 136-713, Korea}  

\abstract {We present  for the first time the nonlinear diffeomorphism in the Randall Sundrum model that can keep the effective Lagrangian invariant in any order of expansion.  We will show that  the off-shell diffeomorphism  shapes  the interaction structure. However the radion mass is  in fact protected by an on-shell  diffeomorphism,  which  can be  spontaneously broken by  the Goldberger-Wise mechanism.  The nonlinear property of diffeomorphism also ensures  a unique radion field even in the extension of  RS model with intermediate branes.}

\begin{document}  
\maketitle 

\section{Introduction}  
Since  the original proposal  of Randall Sundrum (RS) model in the late 90's~\cite{Randall:1999vf, Randall:1999ee},  many attentions are attracted to investigate  the various  phenomenology  in this  warped extra dimension theory.  An interesting aspect is the role  played by spin-2 gravitons when one  embeds the SM particles and dark matters  into the RS model.  As we know,  the  zero mode of spin-2 graviton  precisely reproduces the four dimensional Newtonian gravity.   While  the  massive  Kaluza-Klein graviton can behave as  mediator in the SM particle scattering  by  coupling to  the energy momentum tensor or  be a DM candidate itself~\cite{Babichev:2016bxi, Cai:2021nmk, Falkowski:2020fsu}.   Moreover, self-interactions among the radion and KK gravitons  arise  due to  the nonlinearity.  Although the scattering amplitude of massive  gravitons normally exhibits  highly divergent  behavior in a 4D theory~\cite{Hinterbichler:2011tt},   the unitarity is naturally restored in a UV complete model, like the RS model, due to  the existence of sum rules~\cite{Bonifacio:2019ioc, deGiorgi:2020qlg}. In fact,  the sum rules  originate from   diffeomorphism invariance, that  is  a  local gauge transformation related to the spacetime mapping  and  is the major topic of  this paper.  In principle,  all orders of  expansions from the 5d action should be invariant under  diffeomophism.  At the quadratic order,  the  diffeomorphism  appears  as  a linear gauge  transformation for the graviton and  radion~\cite{Cai:2022geu, Cai:2021mrw}.  In this paper we are going to demonstrate that  the complete gauge transformation should  be altered  to attach with a nonlinear part. Hence the two neighboring orders of Lagrangian expansions are correlated to each other, making the interaction structure  to be governed by  diffeomorphism.  For the 4D  graviton theory in absence of matter field, diffeomorphism invariance  leads to  Bianchi  identity rather than  a conservation law.  The same situation  occurs for the  RS model,  the  off-shell diffeomorphism invariance  is   observed without assuming  the metric fields obey  equation of motions (EOM).    

One important issue is  that the radion  needs to be stabilized,   which requires  its effective  potential obtains a minima at a fixed brane separation. Via the  Goldberger-Wise (GW) mechanism~\cite{Goldberger:1999uk, Goldberger:1999un}, this can be achieved by introducing  a  bulk scalar to couple to the dynamic metric $g_{MN}$.  Without  doubt,  the stabilization breaks down the conformal invariance  at the IR scale from the perspective of  AdS/CFT  correspondence~\cite{Maldacena:1997re, Witten:1998qj, Gubser:1998bc},  hence the continuum  becomes  discrete KK excitations characterized by mass gap.  However the CFT interpretation does not explain  the  origin of  radion's mass.   As we will illustrate,  it is the on-shell diffeomorphism in the fifth dimension  that protects the radion to be massless prior to  stabilization.   Therefore  once the radion acquires its mass, the relevant symmetry  must be broken,  in analogy  to the electroweak gauge symmetry in SM or  global symmetry for  pseudo-Nambu-Goldstone bosons.  This will  lead  to a profound consequence  in a multibrane extension of RS model. In two  papers~\cite{Kogan:2001qx, Pilo:2000et},  the authors claimed that  by  adding a scalar perturbation $ \partial_\mu \partial_\nu f(x,y)$ in the metric  of  a 3-brane RS model,  one can borrow the boundary value of  $f'(x,y)$  to create a second radion, given  that $f(x,y)$  is continuous. However  the derivation  in~\cite{Cai:2022geu} shows that there is no EOM for  $f(x,y)$,  implying it as  a bulk symmetry.   In  fact,  this paper clarifies  that $f(x,y)$ is  an enhanced version of linear diffeomorphism, that does not work in higher orders.  While  the correct nonlinear diffeomorphism  cannot eliminate the $f(x,y)$ perturbation and is on-shell broken after the radion stabilization.  Because the  $f'$ interpretation would  cause  ambiguity in interactions,  the intermediate ``brane''  has  to be rigid  in  the radius stabilization.  

The paper is organized  as follows. In Section 2, after the introduction of  the basic setup in  RS model, we  derive the diffeomorphism transformation rules.  In particular,  we  explicitly show that  at the cubic order,  the  diffeomorphism invariance is  conserved in all interaction vertices, after compensating them with appropriate nonlinear  variation  of  quadratic terms.  In Section 3,  we discuss  the on-shell spontaneous breaking of   diffeomorphism in the fifth dimension due to the GW mechanism.  Finally we  summerize our conclusion in Section 4 and  present the facilitating   details  in  Appendix.

\section{Diffeomorphism in RS model} \label{invariance} 
We start with the five dimensional  action in the RS model, that is composed of  the Einstein-Hilbert action, scalar  and potentials terms:
\beq
S &=&  -  \frac{1}{2 \kappa^2} \int d^5 x \sqrt{g} \, {\cal R} + 
\int d^5 x \sqrt{g}  \Big( \frac{1}{2} g^{IJ}  \partial_I \phi \partial_J \phi
-  V(\phi) \Big)  \nonumber \\ &-&   \int dx^5 \sum_{i} \sqrt{g_4} \lambda_{i}(\phi) \delta(y - y_i) \,, \label{Act} 
\eeq
where  $\kappa^2 = \frac{1}{2M_5^3} \, $ carries the 5d Planck mass $M_5$ and the capital Latin indices $(I,J) = 0, 1, 2, 3, 4 $ run over all the dimensions.  In the RS model,   an $S^1/Z_2$ orbifold  symmetry is imposed  on the $y$-coordinator, with he general metric to be~\cite{Charmousis:1999rg, Csaki:2000zn}:
\beq
d s^2  &=&   e^{-2 A - 2 F }\left[ \eta_{\mu \nu} +  h_{\mu \nu}\right] d x^\mu d x^\nu - \left[1+ G\right]^2 dy^2 \,, \label{metric} 
\eeq  
that  preserves the 4d Poincare  invariance and decouples the graviton  perturbation from the  radion  fields. Note that $g_{\mu 5} =0$ is a direct consequence of  the orbifold symmetry. The Greek indices $(\mu, \nu)$ are restricted   for the  four dimensional Minkowski space.  With  a pure negative cosmological constant $\Lambda$ in the bulk potential, the warped factor in Eq.(\ref{metric}) is derived as  $A(y) = k y, \,  k \equiv \sqrt{- \frac{\Lambda}{12 M_5^3}}$ and the brane tensions are fixed to be $\lambda_i = \pm 12 M_5^3 \,  k $ at the UV and IR branes due to the junction conditions.

The action Eq.(\ref{Act}) includes the GW mechanism used to stabilize the radius of  AdS space  by imposing  appropriate potential in the bulk and branes  to force the bulk scalar develop a $y$-dependent VEV.  The metric  $A(y)$ and the  scalar VEV $\phi_0(y)$ forms a coupled system that is governed by the zeroth order of Einstein equation (See Appendix~\ref{Appendix1}).  

One important aim of this paper is  to elucidate the  symmetries encoded in  Eq.(\ref{Act}).  We will  first investigate  the scenario without adding the GW scalar, i.e. $\phi  =0$. Consider an infinitesimal diffeomorphism,  $X^I \to X^I + \xi^I(X)$,  the  corresponding transformation for the metric is given by its Lie derivatives:
\beq 
\delta g_{IJ} =  \xi^K \, \partial_K  g_{IJ} + \partial_I \xi^K \,  g_{K J} + \partial_J \xi^K \, g_{IK} \,. \label{dG}
\eeq
Under the linear approximation of $g_{IJ}$,  this  will constrain $\xi^K$  to be of  the generic form~\cite{Kogan:2001qx}: 
\beq
\xi^\mu(x,y) &=& \hat{\xi}^\mu(x) +   \partial^\mu    \zeta (x,y)     \,,  \quad    \xi^5(x,y) =  \epsilon(x,y). \label{xi}  
\eeq
with $\epsilon = \zeta' e^{-2A}$ and the prime denoting the fifth dimensional derivative $\partial_5 \equiv \de / \de y$.  This parametrization  will introduce a  perturbation $\partial_\mu \partial_\nu f$ in the metric with a transformation rule of $ f  \to f +  \zeta $.  However as proven in~\cite{Cai:2022geu}, there is no EOM associated with this perturbation, which indicates this parametrization includes redundant degree of freedom.  We can gain more insight  on  this point  by  going  beyond the linear approximation  in order to investigate the nonlinear property like  in General Relativity.  Expanding Eq.(\ref{dG}) till the second order,  the diffeomorphism transformation of $\delta g_{\mu 5}$  is derived as:
\beq
\delta g_{\mu 5} =   \left[ 2 (G +  F) \partial_\mu \epsilon - h_{\mu \nu} \partial^\nu \epsilon \right] = 0    
\eeq 
Since the graviton and radion are independent fields,  we need to set $\partial_\mu \epsilon  = 0$ to avoid the mixing, i.e.  $\xi^5 = \epsilon(y)$ without dependence on $x^\mu$. Then $\partial_\mu \zeta$ as a function of $x^\mu$ alone  will be absorbed into $\hat \xi^\mu$. This property  is intuitively clear,  because  the  gauge parameter associated with the $y$ displacement  should not be correlated to $\hat \xi^\mu(x)$.   With the common factor of $e^{-2A-2F} $ or $(1+G) \eta_{55}$  drops out on both sides of Eq.(\ref{dG}),  one can derive  the complete  transformation rules for all  perturbations in the metric:  
\beq
 \delta h_{\mu \nu} &=&  \left(\partial_{\mu} \hat{\xi}_{\nu} + \partial_\nu \hat{\xi}_\mu  \right)+ \hat{\xi}^\alpha \partial_\alpha  h_{\mu \nu} + \partial_\mu \hat{\xi}^\alpha h_{\alpha \nu} \, \label{hrule}  \\ &+& \partial_\nu \hat{\xi}^\alpha h_{\alpha \mu}  + \epsilon   h'_{\mu \nu}   \nonumber \\
\delta F &=&  A'  \epsilon  +\epsilon F' +\hat{\xi}^\alpha \partial_\alpha F  \, \label{Frule} \\ 
\delta G  &=&  \epsilon' +\partial_5 \left( \epsilon G \right) +\hat{\xi}^\alpha \partial_\alpha G  \,  \label{Grule}
\eeq
Note that the first term in Eq.(\ref{hrule}-\ref{Grule}) is a resemblance of the linear result in~\cite{Cai:2022geu}, with rescaling $\zeta' \to \epsilon \,  e^{2A}$, but differs in $\partial_\mu \epsilon =0$.  While the new  terms  from the nonlinearity  are crucial to comprehend  the  invariance of 5d action at the next to leading order.   It is  a generic fact  that  the diffeomorphism variation of  the metric gives a directional derivative, i.e. $\delta \sqrt{g} = \partial_I (\xi^I \sqrt{g}) = \partial_\mu (\hat \xi^\mu \sqrt{g}) + \partial_5 (\epsilon \sqrt{g})$~\footnote{ In  RS model,  the $\delta_{\epsilon}$ operation on the metric proceeds as follows: $\delta_{\epsilon} \sqrt{g}  =   e^{-4A-4F} \big[ (\delta_{\epsilon} G - 4 (1+ G) \delta_{\epsilon} F) \sqrt{\hat g_4} + (1+ G) \delta_{\epsilon}\sqrt{\hat g_4} \big] =\partial_5 (\epsilon \sqrt{g}) $.  Hence  the cosmology constant term  $\int d^5 x \sqrt g \, \Lambda $  in 5d action is invariant under the diffeomorphism. }. We will recognize below  the gauge symmetries generated by $\hat \xi^\mu$ and $\epsilon$ ensure the zero mode of  graviton and  the radion to be  massless respectively.  
 
First of all, we are going to illustrate that prior to  radion stailization  how the gauge transformations Eq.(\ref{hrule}-\ref{Grule}) operate in Eq.(\ref{Act})  with $V = \Lambda$ and $\phi =0$. For convenience, one can define  the 4D effective Lagrangian as $\mathcal{L} = -\frac{1}{2 \kappa^2}\int dy \sqrt{g} \left( {\cal R}  + \Lambda \right)$, without considering the brane terms. Under the diffeomorphism,  the total variation of  bulk action  is  of the following form: 
\beq
\delta S =  \int d^4 x   \left[ \frac{  \partial \mathcal{L}}{ \partial F} \delta F  +  \frac{  \partial \mathcal{L}}{ \partial G} \delta G +  \frac{  \partial \mathcal{L}}{ \partial h_{\mu \nu}} \delta h_{\mu \nu }\right]  
\eeq  
In the following, we will use a superscript to mark  the linear and nonlinear parts of variations: $\delta = \delta^{(1)} + \delta^{(2)}$. To declare diffeomorphism as a continuous symmetry,  the transformation rules in Eq.(\ref{hrule}-\ref{Grule}) should keep the 5d action invariant up to a surface term.  Schematically,   in terms of metric  perturbation expansion $\mathcal{L} = \sum_n \mathcal{L}^{(n)}$,  the diffeomorphism will require $ \int d^4 x \left[ \delta^{(2)} \mathcal{L}^{(n-1)} + \delta^{(1)}  \mathcal{L}^{(n )} \right]  = 0$.
Before going beyond,  we recall  that  the  effective Lagrangian  at the quadratic  order is~\cite{Cai:2022geu}:
\beq     
\mathcal{L}^{(2)} & = &  - \frac{1}{2\kappa^2}  \int dy e^{-2A}  \Big\{  {\Big[ \tilde{\mathcal L}_{FP} } + \frac{e^{-2A}}{4} \left(\Big[ h'_{\mu \nu} \,  h'^{\mu \nu} - h'^2 \Big] + 6 \Big[ h_{\mu \nu}  h^{\mu \nu} - \frac{1}{2} h^2 \Big] A''  \right) \Big] 
\nonumber \\  &+&  6  \Big[ \partial_\mu F \partial^{\mu} (F-G)  
-  2  \, e^{-2A}  \left(  \left[ F'-  A' G  \right]^2  + 4 F^2 A'' \right)  \Big]  + \mathcal{L}_{mix}\Big\}  
\label{quad}     
\eeq
where $\tilde{\mathcal{L}}_{FP}$ denotes the primitive  Fierz-Pauli Lagrangian reported in  Eq.(\ref{FP}).   
After integration by parts, it  transforms into  the  familiar form:
\beq
{\mathcal L}_{FP}= \frac{1}{2} \partial_\nu h_{\mu \alpha} \, \partial^\alpha    
h^{\mu \nu} -   \frac{1}{4} \partial_\mu h_{\alpha \beta} \, \partial^\mu h^{\alpha \beta}     
- \frac{1}{2} \partial_\alpha h \, \partial_\beta h^{\alpha \beta} +\frac{1}{4}    
\partial_\alpha h \, \partial^\alpha h   \label{FP2}
\eeq
And the direct expansion also includes a mixing term: 
\beq
 \mathcal{L}_{mix} = \left( G - 2F\right) \left[ \partial_\mu \partial_\nu h^{\mu \nu} -\Box h \right] + 3  e^{-2A} \left[ \left( F' - A' G \right) h'  + 4 F h A'' \right]  \label{Lmix}
\eeq
The $A''$ terms in Eq.(\ref{quad}) and Eq.(\ref{Lmix}) will be cancelled by the same expansions from  the $\lambda_i$ brane action. As the tadpole comprises only  kinetic terms before stabilization~\cite{Cai:2021mrw},   the quadratic action $\int d^4 x \, \mathcal{L}^{(2)}$ is invariant  by itself under the linear  transformation.  It is worthwhile to mention that as  the mixing terms  are  necessary to prove  the diffeomorphism invariance in an off-shell approach,  we will not impose the condition: $\partial_\mu h^{\mu \nu} = h =0$.  
Note that  a 5d mass term $m^2 \left[h^2 - h^{\mu \nu} h_{\mu \nu}\right]$ of graviton is forbidden by $\hat \xi^\mu$ transformation\footnote{However the mass of KK gravitons, except for the zero mode with a constant 5d profile, can be generated  from the term $\propto  \left(h'^{\mu \nu} h'_{\mu \nu} -h'^2\right)$  in Eq.(\ref{quad}) that is permitted by the $\delta_{\xi}$ diffeomorphism. },  and the  radion  remains massless  because  $F' - A' G = 0$  is a conserved quantity under  Eq.(\ref{Frule}-\ref{Grule}).

The cubic order expansions of 4D effective Lagrangian without the GW stabilization are reported in Appendix~{\ref{Appendix2}}.  Inherited from the characteristics of Ricci scalar, these terms are second order in derivatives and  divided into two categories:  one involving  two 4D derivatives and the other  with  two $\partial_5$ derivatives.  Now  diffeomorphism invariance will force the linear transformation  of  cubic  terms  to match with the nonlinear  one  of quadratic terms, in line with the  constraint  $ \int d^4 x \left[ \delta^{(2)} \mathcal{L}^{(2)} + \delta^{(1)}  \mathcal{L}^{(3)} \right]  = 0 $.  Indeed, at the cubic order,  the gauge invariance of each interaction  type is guaranteed  independently.   As a result,  the  diffeomorphism  plays a significant role in shaping the coupling structure. 
\begin{itemize}
\item Vertices with 3 radions  

We will first study the case of 3 radion coupling.  Ignoring the term proportional to  $A'' = \frac{\kappa^2}{3} \sum_i \lambda_i \delta(y-y_i)$ in Eq.(\ref{3F}), the  Lagrangian of our concern  can split into two parts $\mathcal{L}_{F^3+ F^2G}=\mathcal{L}_{\partial_\mu^2 F} + \mathcal{L}_{\partial_5^2 F}$, that reads:
\beq
\mathcal{L}_{\partial_\mu^2 F^3}&=&  - \frac{3}{ \kappa^2} \int  dy \, e^{-2A }   \Big[ G \partial_\mu F \partial^\mu F - 2 F  \partial_\mu F \partial^\mu \left(F-G\right) \Big]  \\
 \mathcal{L}_{\partial_5^2 F^3}   &= &   - \frac{1}{ \kappa^2} \int dy \, e^{-4A}   \Big[ 6 \left( F' - A' G \right)^2 (G+ 4 F) \nonumber \\
&+& \frac{32}{3}A' \left( 3 \left(F' - A' G \right) F^2   +  F^2 \left( 3G - 4F\right)  A' \right) \Big]  \, \label{L3F}
\eeq
Note that since the transformation rules in Eq.(\ref{hrule}-\ref{Grule}) do not convert $\partial_\mu$ to $\partial_5$, we can separately examine the variations on  two equations above. Under the linear transformation related to the gauge parameter $\epsilon$,  one can find that the variation of 3 radion vertices with two $\partial_\mu$ is:
 \beq
  \delta_{\epsilon}^{(1)} \mathcal{L}_{\partial_\mu^2 F^3}  = -\frac{3}{\kappa^2}  \int dy \Big[ \partial_5 (e^{-2A} \epsilon )
  \partial_\mu F \partial^\mu F + 2 e^{-2A} \epsilon A'  \partial_\mu F \partial^\mu G  \Big] \, \label{d3F0}
 \eeq
 and the nonlinear transformation for the radion kinetic term is of the same form: 
 \beq
&&  \delta_{\epsilon}^{(2)}  \left(  \frac{3}{ \kappa^2} \int dy e^{-2A}  \partial_\mu F \partial^\mu (G-F)  \right)
 \nonumber \\ &=&  -\frac{3}{\kappa^2} \int dy e^{-2A} \epsilon  \Big[ \partial_5
 ( \partial_\mu F \partial^\mu F) - 2 A'  \partial_\mu F \partial^\mu G \Big]  \label{d2F0}
 \eeq
Hence the summation of  Eq.(\ref{d3F0}) and Eq.(\ref{d2F0}) vanishes identically: 
\beq
 \delta_{\epsilon}^{(2)} \left( \frac{3}{ \kappa^2} \int dy e^{-2A}   \partial_\mu F \partial^\mu (G-F)  \right) + \delta_{\epsilon}^{(1)} \mathcal{L}_{\partial_\mu^2 F^3}  = 0 \label{d3Ftot}
\eeq
 which indicates that  $\mathcal{L}_{\partial_\mu^2 F^3}$  combined with the radion kinetic term is invariant under the $\delta_\epsilon$ diffeomorphism.
 Note that  in  deriving Eq.(\ref{d3Ftot}),  we do not  require any equation of motion  at all steps. And the invariance is satisfied in an off-shell manner. Then we can figure out what happens to  the part with two $\partial_5$.  After  operating $\delta_\epsilon$ on  Eq.(\ref{L3F}),  we will obtain that:
\beq
\delta_\epsilon^{(1)} \mathcal{L}_{\partial_5^2 F^3} =  -\frac{6}{ \kappa^2} \int dy  e^{-4A} \left[ \left( F'- A'G \right)^2 \left[  \epsilon' + 4 A' \epsilon  \right]  + A'' \mbox{terms} \right] \,  \label{d3F1}
\eeq 
where the variation of  the second line in Eq.(\ref{L3F}) only contributes to a $A''$ term.  And for the mass term in Eq.(\ref{quad}),  the nonlinear  $\delta_\epsilon$  variation gives:
\beq
&& \delta_\epsilon^{(2)} \left( \frac{6}{\kappa^2}  \int dy e^{-4A}  \Big[ F'-  A' G  \Big]^2 \right) \nonumber \\
&=&   \frac{12}{\kappa^2}  \int dy e^{-4A}  \left( F' -A' G\right) \Big[ \partial_5 \left(\epsilon  F' \right) - A' \partial_5 \left( \epsilon  G\right)\Big] \label{d2F1}
\eeq
The combination of  Eq.(\ref{d3F1}-\ref{d2F1})  yields a total derivative  plus  $A''$ surface terms that vanish upon the $y$ integration for $\epsilon =0$ at the branes.  Therefore,  the off-shell diffeomorphism is fully  preserved for all terms in   Eq.(\ref{3F}).

\item Vertices with  one graviton and two radions 

The cubic interaction with a single graviton is described  by Eq.(\ref{h2F}).  We will defer the discussion of  terms containing two  $\partial_5$.  Then the effective Lagrangian  reduces to be:
 \beq
 \mathcal{L}_{\partial_\mu^2 hF^2} = &-& \frac{1}{2 \kappa^2} \int dy e^{-2A}  \Big[  2 F (G-F) \left( \Box h - \partial_\mu \partial_\nu h^{\mu \nu} \right)  \nonumber \\ &+&   6 h^{\mu \nu }  \partial_\mu F \partial_\nu (G- F)  - 3 \partial_\mu F \partial^\mu (G-F) h  \Big]   \, \label{Lh2F}
 \eeq
For the  $\delta_{\hat \xi}$ diffeomorphism,  the first term  in Eq.(\ref{Lh2F})  is invariant under  the leading order transformation.  While  the linear  $\delta_{\hat \xi}^{(1)}$ variation on the remaining terms  is:
 \beq
  \delta_{\hat \xi}^{(1)}  \mathcal{L}_{\partial_\mu^2 hF^2}  &=& - \frac{3}{\kappa^2} \int dy \, e^{-2A}\Big[ \left(\partial^ \mu \hat \xi^\nu +  \partial^ \nu \hat \xi^\mu \right) \partial_\mu  F \partial_\nu (G-F) \nonumber \\ &-&  \left(  \partial_\nu \hat \xi^\nu \right)  \partial_\mu  F  \partial^ \mu (G-F)\Big]   \label{dh2F}
  \eeq 
Then we  accompany  the above ansatz with  the nonlinear $\delta_{\hat \xi}^{(2)}$ variation on the radion kinetic term: 
\beq  
\delta_{\hat \xi}^{(2)}   \left( \frac{3}{ \kappa^2} \int dy e^{-2A}  \partial_\mu F \partial^\mu (G-F)  \right) 
 &=& \frac{3}{\kappa^2} \int dy \, e^{-2A} \Big[   \hat \xi^\nu \partial_\nu \Big(  \partial_\mu  F  \partial^ \mu (G-F) \Big)   \nonumber \\ &+& \left(\partial^ \mu \hat \xi^\nu +  \partial^ \nu \hat \xi^\mu \right) \partial_\mu  F \partial_\nu (G-F)   \Big]
 \eeq
  where  the term in the second line  exactly nullifies the first line in  Eq.(\ref{dh2F}). Hence the net effect  becomes:
  \beq
  &&  \delta_{\hat \xi}^{(2)}   \left(  \frac{3}{ \kappa^2} \int dy e^{-2A}  \partial_\mu F \partial^\mu (G-F)  \right)  +  \delta_{\hat \xi}^{(1)}   \mathcal{L}_{\partial_\mu^2 hF^2}  \nonumber \\
  &=& \frac{3}{\kappa^2} \int dy \, e^{-2A} \partial_\nu \Big[  \hat \xi^\nu  \partial_\mu  F  \partial^ \mu (G-F)  \Big]
 \eeq
Due to $\hat \xi^\mu = 0$ at the boundary,   the surface term vanishes when subjected to 4-dimensional integration, in accordance with the symmetry.
Subsequently  we can establish  the $\delta_{\epsilon}$ diffeomorphism for Eq.(\ref{Lh2F}),  where only the first term matters owing  to  $\partial_\mu \epsilon =0$.
Remarkably,  its linear $\delta_{\epsilon}$ variation  aligns with the nonlinear $\delta_{\epsilon}$ variation of a mixing term  in Eq.(\ref{Lmix}): 
\beq
\delta_{\epsilon}^{(2)} \left ( \frac{1}{2 \kappa^2}\int dy e^{-2A}  \left( G-2 F \right) \left[\Box h -  \partial_\mu \partial_\nu h^{\mu \nu}  \right] \right) +  \delta_{\epsilon}^{(1)} \mathcal{L}_{\partial_\mu^2 hF^2} =0 
\eeq 
Finally  we  comment on the  $h$ and $h'$ terms in Eq.(\ref{h2F}). They  can  be generated  from the mass term $ \propto (F'-A' G)^2$ and the mixing term $ \propto (F'-A' G) h'$ respectively via  the nonlinear gauge transformation.

\item Vertices with two gravitons and one radion    

We proceed to examine the cubic  interaction with two gravitons  and  the relevant Lagrangian is:
 \beq
 \mathcal{L}_{h^2 F+ h^2 G} &=&   \frac{1}{2 \kappa^2}\int dy \,  e^{-2A} \Big\{ \left(2F- G \right) \tilde{\mathcal{L}}_{FP} 
 +   e^{-2A} \Big[ \frac{1}{4} \left(G+4F \right) \big[ h'_{\mu \nu} h'^{\mu \nu} - h'^2  \big]  \nonumber \\&+& 3 \left( F' - A' G \right) \big( h'_{\mu \nu} h^{\mu \nu} -\frac{1}{2} h' h \big) \Big] \Big\} \, \label{L2hF}
 \eeq
 where $\tilde{\mathcal{L}}_{FP}$ can not be performed  partial integration and  the $A''$ term in Eq.(\ref{2hF}) is dropped.  Note that for a stabilized RS,  the terms  proportional to  $\left(F'- A'G \right)$  and $A''$  have significance and their effects will  be addressed  in Appendix~\ref{Appendix3}.  In the two-graviton case, we initially demonstrate the  $\delta_{\epsilon}$ invariance of Eq.(\ref{L2hF}). 
 \beq
 \delta_{\epsilon}^{(1)} \mathcal{L}_{h^2 F+ h^2 G} 
= \frac{1}{2 \kappa^2}\int dy \Big[ \frac{e^{-4A}}{4} \left( \epsilon' + 4 A' \epsilon  \right) \left[ h'_{\mu \nu} h'^{\mu \nu} - h'^2 \right] - \partial_5 \left(\epsilon  e^{-2A}\right) \tilde{\mathcal{L}}_{FP} \Big] \label{d2hF}
 \eeq
 The same pattern exists for the  nonlinear $\delta_{\epsilon}$ variation on the quadratic graviton terms in Eq.(\ref{quad}) after the partial integration:
\beq
&&  \delta_\epsilon^{(2)} \left( - \frac{1}{2 \kappa^2} \int dy e^{-2A} \Big[ \tilde{\mathcal{L}}_{FP} + \frac{e^{-2A}}{4} \left[ h'_{\mu \nu} h'^{\mu \nu} - h'^2 \right] \Big] \right) \nonumber \\
&=& - \frac{1}{2 \kappa^2} \int dy \Big[ \epsilon e^{-2A} \,  \partial_5 \tilde{\mathcal{L}}_{FP}   +  \frac{e^{-4A}}{4} \Big( \epsilon'  + 4 A' \epsilon  \Big)  \Big[ h'_{\mu \nu} h'^{\mu \nu} - h'^2 \Big] \Big] 
 \eeq
which  offsets  with Eq.(\ref{d2hF}), hence  the  $\delta_{\epsilon}$ invariance is evident.   As anticipated,  the  $\delta_{\hat \xi}$ diffeomorphism is similarly manifest in an off-shell style  for this interaction. The second term  in Eq.(\ref{L2hF}) is invariant  under  the linear order transformation, 
while  the linear transformation  of  first and third terms  precisely cancel the nonlinear variation of corresponding quadratic terms in Eq.(\ref{quad}):
 \beq
\frac{1}{2 \kappa^2}\int d^5 x \,  e^{-2A}  \Big[\delta_{\hat \xi}^{(2)} \left( ( 2F- G) [ \partial_\mu \partial_\nu h^{\mu \nu} -\Box h ] \right) + (2F-G) \delta_{\hat \xi}^{(1)} \tilde{\mathcal{L}}_{FP}  \Big]= 0 \, \label{dF2h}
 \eeq
 \beq
 - \frac{3}{2 \kappa^2}\int d^5 x  e^{-4A} \Big[ \delta_{\hat \xi}^{(2)} \left( (F' - A' G ) h' \right)-  ( F' - A' G) \delta_{\hat \xi}^{(1)} \big( h'_{\mu \nu} h^{\mu \nu} -\frac{1}{2} h' h \big)\Big]  =0 
\eeq

\item Vertices with three gravitons  
 
We  briefly investigate the structure of  3-graviton interaction,  excluding  the  part  that is  identical to those in  General relativity. The cubic Lagrangian with two $\partial_5$ is:
 \beq
 \mathcal{L}_{h^3} &=&   \frac{1}{4 \kappa^2}\int dy \,  e^{-4A} \Big[ h^{\mu \nu} \left( h'_{\mu \alpha} h'^\alpha_\nu  - h'_{\mu \nu} h'  \right)  + \frac{1}{4} h \left( h'^2 - h'^{\mu \nu} h'_{\mu \nu} \right)  \Big] \, \label{L3h}
 \eeq
 which along with the quadratic term $\propto  \left[  h'_{\mu \nu} h'^{\mu \nu} - h'^2  \right]   $  satisfies the $\delta_{\hat \xi}$ diffeomorphism:
 \beq
 \delta_{\hat \xi}^{(2)} \left(  - \frac{1}{8 \kappa^2} \int d^5 x   e^{-4A}  \left[  h'_{\mu \nu} h'^{\mu \nu} - h'^2  \right] \right) + \delta_{\hat \xi}^{(1)}   \int d^4 x  \, \mathcal{L}_{h^3} = 0
  \eeq
Thus  we have demonstrated  the off-shell diffeomorphism for all types of interactions at the cubic order. 
\end{itemize}

\section{On-shell spontaneous symmetry breaking} \label{5d-breaking} 
After the GW scalar develops a VEV,  the back reaction on the metric  will make  the radion massive.  One would  expect that the  relevant symmetry is spontaneously broken by the GW mechanism in a context with  impact  merely in  the radion sector.  However since  the diffeomorphism is an off-shell symmetry,   its breaking can only occur  by imposing conditions from equations of motion (EOM),  which makes it an on-shell breaking phenomenon.  For the off-shell symmetry,  it is necessary to amend a transformation rule for the GW  scalar with a $y$-dependent VEV: 
\beq
\delta \varphi = \epsilon  \phi'_0  + \epsilon  \varphi' + \hat \xi^\alpha \partial_\alpha \varphi  \label{Vrule} \,.
\eeq
Note that Eq.(\ref{Vrule}) along with Eq.(\ref{hrule}-\ref{Grule})  forms a full set of diffeomorphism transformation, that  ensures the invariance of  the complete 5d action in  Eq.(\ref{Act}). 
As a simple verification, one can confirm that this transformation keeps  the following 4D derivative terms,
\beq
 \int dy  \, e^{-2A} \Big[ \left(1 + G- 2F + \frac{1}{2} h \right) \partial^\mu \varphi \partial_\mu \varphi - h^{\mu \nu}\partial_\mu \varphi \partial_\nu \varphi \Big] \,,
\eeq
 invariant up to the cubic order. But  the off-shell symmetry is  broken when we impose on the 5d action with  the $(\mu 5)$-component Einstein equation for  a massive radion~\cite{Cai:2021mrw, Csaki:2000zn}:
\beq
 F'  - A' G  = \frac{\kappa^2}{3} \phi'_0 \varphi    \,.  \label{gauge}    
\eeq  
To shed light on  this point, one can  examine the variation of Eq.(\ref{gauge}) under Eq.(\ref{Frule}-\ref{Grule}) and  Eq.(\ref{Vrule}). Although Eq.(\ref{gauge}) remains invariant up to a boundary term under the linear transformation,  this property is violated by  the  nonlinear $\delta_{\epsilon}$ transformation except for $\phi'_0 =0 $.   Hence we can consider the quality of $F' - A' G $  as a symmetry  index. Once this index becomes nonzero, the $\delta_{\epsilon}$ symmetry  is on-shell broken, leading to the mass acquisition  by the radion.

In the following discussion, we will demonstrate the evident on-shell symmetry breaking in the potential sector of the radion field. Note that, after radion stabilization only the kinetic part survives in the tadpole term $\mathcal{L}_{tad}$\footnote{From the linear expansion in \cite{Cai:2021mrw}, one can calculate  $ \mathcal{L}_{tad}  = - \frac{1}{\kappa^2} \int  dy e^{-2 A} \Box \left( 3F - G\right)  \xRightarrow[]{G= 2F} \frac{m^2}{\kappa^2} \int dy e^{-2A} F$,  that is the same as the result derived in~\cite{Cai:2022geu}, where   Einstein equations are applied to transform  the Ricci scalar into the trace of energy momentum tensor.}, whereas the potential part, involving $\partial_5$ derivatives, vanishes upon enforcing Eq.(\ref{BG0}-\ref{BG2}).  Consequently, we initiate our analysis of the effective potential at the quadratic order within a stabilized RS model. The  $\partial_5$ term in Eq.(\ref{quad}) needs to be augmented with quadratic expansions of $\varphi$  from Eq.(\ref{Act}):  
\beq
  \mathcal{L}_m & = &  -\frac{1}{2} \int dy e^{-4A}    \Big\{ - \frac{12}{\kappa^2} \big[ F' -   A' G  \big]^2  +  \varphi'^2 + G^2  \phi'^2_0  - 2 (G+ 4F) \phi'_0 \varphi' \nonumber \\ & +& \left[ 2 (G- 4F)  \frac{\partial V}{\partial \phi_0} \varphi  +  \frac{\partial^2 V}{\partial \phi_0^2} \varphi^2 \right]  -  \sum_i \left[  8  \frac{\partial \lambda_i}{ \partial \phi_0} F \varphi  -  \frac{\partial^2 \lambda_i}{\partial \phi_0^2}\varphi^2  \right] \delta(y-y_i)  \Big\} \, \label{mass}
\eeq
The linear $\delta_{\epsilon}$ variation of  Eq.(\ref{mass}) yields: 
\beq
\delta_{\epsilon}^{(1)} \mathcal{L}_m &=& -  \int dy e^{-4A}    \Big\{  - \frac{12}{\kappa^2} \Big[ \left( F' - A' G  \right) \epsilon A''  \Big]  +  \varphi'  \epsilon  \left( \phi''_0 - 4A' \phi'_0 \right)  -  4 \phi'_0 F \partial_5 \left( \phi'_0 \epsilon  \right)  \nonumber \\ & -  &  \phi'_0 G  \phi''_0 \epsilon   +  \frac{\partial V}{\partial \phi_0}   \left(G- 4F\right) \phi'_0 \epsilon +    \frac{\partial V}{\partial \phi_0}  \varphi \left[ \epsilon'  - 4 A' \epsilon \right]  +  \frac{\partial^2 V}{\partial \phi^2_0}  \varphi \phi'_0 \epsilon  \Big\}  \label{dL2}
\eeq
The combination of some terms  in Eq.(\ref{dL2}) gives a surface term:
\beq
&&  \int dy e^{-4A}   \left[    \varphi'  \epsilon  \left( \phi''_0 - 4A' \phi'_0 \right)+   \frac{\partial V}{\partial \phi_0}  \varphi \left( \epsilon'  - 4 A' \epsilon \right)  +  \frac{\partial^2 V}{\partial \phi^2_0}  \varphi \phi'_0 \epsilon \right] \nonumber \\
&=&   \int dy \left[ e^{-4A}  \varphi' \epsilon \sum_i \frac{\partial \lambda_i}{\partial \phi_0} \delta(y- y_i) + \partial_5 \left(   \frac{\partial V}{\partial \phi_0}  \varphi \, \epsilon \, e^{-4A}\right) \right]    \label{reform1} 
\eeq
where we apply  Eq.(\ref{BG0}). Then performing partial integration to the last term in the first line of Eq.(\ref{dL2})  and vanishing the terms in Eq.(\ref{reform1}), one can convert Eq.(\ref{dL2}) to a new form:   
\beq
\delta_{\epsilon}^{(1)} \mathcal{L}_m &=&  \frac{12}{\kappa^2} \int dy     e^{-4A}    \Big\{ \Big[  \left( F' - A' G  \right) \epsilon  A''  \Big]  -   \frac{\kappa^2}{3} \phi'^2_0 \epsilon \left(F'-A' G \right) \Big\}
\eeq
which is also a surface term due to Eq.(\ref{BG1}). With  lengthy algebras,  we derive  that  $\mathcal{L}_m$ is invariant under the linear $\delta_{\epsilon}$ diffeomorphism.

The on-shell  symmetry breaking effects  begin to emerge beyond the quadratic order, as they are inherently tied to the nonlinear $\delta_\epsilon$ variation.   Expanding the effective potential  till  the cubic order gives:
\beq
 \mathcal{L}_{\rm cub} &=&   \mathcal{L}_{\partial_5^2 F^3} - \frac{1}{2} \int dy e^{-4A} \Big[  2 \phi'_0 \varphi' \left(G^2 +4 FG + 8 F^2\right) + \Big( \frac{64}{9}  F^3 -  G^2 ( G + 4 F) \Big)\phi'^2_0 \nonumber \\
& - & (G+ 4F) \varphi'^2  -  8F(G-2F) \frac{\partial V}{\partial \phi_0} \varphi +  (G-4F) \frac{\partial^2 V}{\partial \phi_0^2} \varphi^2
+ \frac{1}{3}  \frac{\partial^3 V}{\partial \phi_0^3} \varphi^3 \Big] 
\eeq
where  $ \mathcal{L}_{\partial_5^2 F^3} $ contains the terms proportional to  $F'-A'G$. Next we can verify  whether the nonlinear transformation of $\delta_{\epsilon}^{(2)} \mathcal{L}_m$  aligns with the linear transformation of $ \delta_{\epsilon}^{(1)} \mathcal{L}_{\rm cub}$  after imposing  the condition of Eq.(\ref{gauge}). With the similar calculation, we find that:
\beq
&&\delta_{\epsilon}^{(2)}  \left(\mathcal{L}_m|_{F'- A' G = \frac{\kappa^2}{3} \phi'_0 \varphi}  \right) + \delta_{\epsilon}^{(1)} \left( \mathcal{L}_{\rm cub}|_{F'- A' G = \frac{\kappa^2}{3} \phi'_0 \varphi}  \right) \nonumber \\  &=& \frac{2 \kappa^2}{3} \int dy e^{-4A} \phi'_0 \varphi \left( \phi^{\prime 2}_0 G \epsilon + \phi''_0 \varphi \epsilon + \phi'_0 \varphi \epsilon' \right)
\eeq
that  is not vanishing  because of $\phi'_0 \neq 0$.  This  implies that  the $\delta_\epsilon$ diffeomorphism is indeed on-shell broken when the 5d action is subject to the EOM  constraint given by  Eq.(\ref{gauge}).

\section{Conclusion}  
In this paper, we work out the 4D  effective Lagrangian till the cubic order  and  elucidate the symmetries that shape the interaction structures  in  the  RS model. Following General Relativity, we derive  the nonlinear diffeomorphism transformation rules Eq.(\ref{hrule}-\ref{Grule})  for the metric outlined in Eq.(\ref{metric}).  Our derivation reveals that  all interactions  preserve an off-shell  diffeomorphism.   It is  demonstrated that  the GW mechanism  spontaneously breaks the $\delta_{\epsilon}$ diffeomorphism in an on-shell approach  to make the radion massive. An interesting observation, as presented in Appendix C, is that the cubic couplings with two gravitons and one radion adhere to the on-shell diffeomorphism after GW stabilization.  Consequently,  the sum rules that restore the unitarity of  graviton pair scattering will  remain valid in a stabilized RS model.

The nonlinearity is  naturally inherited in  the Einstein-Hilbert action. This work clarifies  that  the gauge parameter  $\epsilon(y)$ in  5d  diffeomorphism should not depend on the coordinate $x^\mu$, in order to eliminate the nonlinear order terms in   $\delta g_{\mu 5}$.  This property confirms that there is only  a unique radion even in the extensions of  RS model with multiple branes.  To interpret  this point, let us suppose to add  a  perturbation of the form $\partial_\mu \partial_\nu f(x, y)$ in the 4D metric $g_{\mu \nu}$.   A disguising result is that at the quadratic order expansion of 5d action, the $ f(x, y)$ perturbation without EOM only shows up in the $\mathcal{L}_{mix}$, and  can be fully removed by  the gauge fixings~\cite{Cai:2022geu}.  While  the terms with four $\partial_\mu$ derivatives  will appear in the cubic order,  when one substitutes   $h_{\mu \nu} \to \partial_\mu \partial_\nu f(x, y)$ in Eq.(\ref{Lh2F}) and Eq.(\ref{L2hF}).  To avoid  ambiguity arising from these terms,  a  rigorous  symmetry is needed  to remove $f(x,y)$,  while simultaneously  keep the other terms invariant.  The diffeomorphism surely satisfies  the latter condition.  However within our gauge symmetry,  the  $f(x,y)$ terms  are  also  left to be  invariant  under the transformation, such that  ambiguity is inevitable.  Since there is no radion associated with the rigid intermediate brane,  the stabilization of 3-brane RS model  should  be carried out using  the approach in \cite{Cai:2021mrw}.

\section*{Note Added} 
 A recent paper~\cite{Lee:2021wau} considers two radions in a 3-brane RS model by using  the  $\partial_\mu \partial_\nu f (x,y)$ perturbation in the metric.  That scheme requires the truncation of 5d action till the quadratic order. In fact,  the trick to create a second radion  with  the boundary value of $f'(x,y)$ (equivalent to a gauge fixing)  will cause ambiguity in the interaction terms, even under the circumstance of not  stabilizing the radions~\cite{Kogan:2001qx}. One more issue is that the authors of~\cite{Lee:2021wau} constrain  each KK excitation in the radion-scalar system  to only reside  in  one subregion.  This violates the boundary conditions in  that paper's Eq.(3.25-3.26). The KK excitations should be treated on an equal footing like the stabilized zero modes, because  they hold exactly the same EOM and boundary conditions.  

\section*{Acknowledgments}
H.C. \ is supported by the National Research Foundation of Korea (NRF) grant funded by the Korea government (MEST) (No. NRF-2021R1A2C1005615).

\newpage

\input{appendix}

\newpage
\bibliographystyle{JHEP}
\bibliography{symmetry}

\end{document}

%% file: appendix.tex
\appendix    

\section{Metric background solution}~\label{Appendix1}
The coupled equations of background metric are  derived using the linearized Einstein equation: $ R_{MN}  = \kappa^2 \left(T_{MN} -\frac{1}{3} \, g_{MN} T^a_a\right)$:  
\beq
&&\phi_0 '' = 4 A' \phi_0 ' + \frac{\partial V(\phi_0)}{\partial \phi}+ 
\sum_{i} \frac{\partial \lambda_i(\phi_0)}{\partial \phi} \delta(y-y_i) \,,  \label{BG0} \\
&& A'' = \frac{ \kappa^2 }{3} \phi'^2_0 + \frac{\kappa^2 }{3}
\sum_i \lambda_i(\phi_0) \delta(y-y_i) \,, \label{BG1}
\\
&& A'^2 = \frac{\kappa^2 {\phi_0}'^2}{12}  
- \frac{\kappa^2}{6} V(\phi_0) \,.
\label{BG2} 
\eeq   
For a specific form of  $V(\phi_0) $, general  solutions for those equations  can be written in terms of  the superpotential~\cite{DeWolfe:1999cp, Behrndt:1999kz}.

\section{Effective Lagrangian at the cubic order}~\label{Appendix2} 
In a RS model  without radion stabilization,  the 5d action is: 
\beq
S  = -\frac{1}{2 \kappa^2}\int dx^5 \sqrt{g} \left( {\cal R}  + \Lambda \right) -    \int dx^5 \sum_{i}  \sqrt{g_4}  \lambda_i \delta(y - y_i). \label{AdS}
\eeq
The 4D  effective Lagrangian in the bulk  is defined  to be:
\beq
 \mathcal{L}_{eff} = -\frac{1}{2 \kappa^2}\int dy \sqrt{g} \left( {\cal R}  + \Lambda \right)  \label{eff}
\eeq
Expanding Eq.(\ref{eff}) till the third order,  the Lagrangian for cubic interactions except for the 3-graviton vertices  can be collected as:
\beq
 \mathcal{L}_{F^3+ F^2 G} = & - & \frac{1}{\kappa^2}\int dy e^{-4A}  \Big\{ 3 e^{2A}   \Big[ G \partial_\mu F \partial^\mu F - 2 F  \partial_\mu F \partial^\mu \left(F-G\right) \Big] \nonumber \\
  &+& 2 \left(F' - A' G \right) \Big[ 3(4 F + G ) \left( F' - A' G \right) +16 F^2 A' \Big] \nonumber \\
 &+&  \frac{32}{3} F^2 \Big[ A'^2 (3 G-4 F)  + 4 F A'' \Big] \Big\}     \label{3F} 
 \eeq  
 
 \beq
 \mathcal{L}_{h F^2+ h FG} = &-& \frac{1}{2 \kappa^2}\int dy e^{-4A}  \Big\{ e^{2A} \Big[  2 F (G-F) \Box h  - 3 \partial_\mu F \partial^\mu (G-F) h  \nonumber \\ &+&   6 h^{\mu \nu }  \partial_\mu F \partial_\nu (G- F) - 2 \partial_\mu \partial_\nu h^{\mu \nu}  F (G-F)   \Big]  -24  F^2  h A'' \nonumber \\
& - & 3 \left(F' - A' G \right) \Big[2 h  \left(F'-A' G \right) +  h' (G+ 4 F)  \Big]  \Big\}  \label{h2F} 
 \eeq

\beq
 \mathcal{L}_{h^2 F+ h^2 G} &=&  \frac{1}{2 \kappa^2}\int dy e^{-4A} \Big\{ e^{2A} \left(2F-G\right) \tilde{\mathcal{L}}_{FP} +  \Big[ \frac{1}{4} \left(G+4F \right) \left( h'_{\mu \nu} h'^{\mu \nu} -h'^2  \right) \nonumber \\&+& 3 \left( F' - A' G \right) \Big( h'_{\mu \nu} h^{\mu \nu} -\frac{1}{2} h' h \Big) + 6 F \Big(h_{\mu \nu} h^{\mu \nu} -\frac{1}{2} h^2 \Big)A'' \Big]  \Big\} \label{2hF}
 \eeq 
 where  $\tilde{\mathcal{L}}_{FP}$ is  the untransformed  Fierz-Pauli Lagrangian  derived from  the  equations in  Appendix (A.1) of Ref.{\cite{Cai:2022geu}}, with:
 \beq
 \tilde{\mathcal{L}}_{FP} &=& h^{\mu \nu} \Box h_{\mu \nu} + \frac{3}{4} \partial^\alpha h^{\mu \nu} \partial_\alpha h_{\mu \nu}  - \Big( 2 h^{\mu \nu}  \partial_\mu \partial_\alpha h^\alpha_\nu + \partial^\mu h_{\mu \alpha} \partial^\nu h^\alpha_\nu + \frac{1}{2} \partial^\nu h _{\mu \alpha} \partial^\mu h^\alpha_\nu \Big) \nonumber \\
 &+& \partial^\mu h_{\mu \nu} \partial^\nu h + h_{\mu \nu} \partial^\mu \partial^\nu h + \frac{1}{2} h  \partial^\mu \partial^\nu  h_{\mu \nu} - \frac{1}{4} \left( \partial_\mu h \partial^\mu h  +2  h \Box h \right) \, \label{FP}
 \eeq 
Note that our cubic couplings  are  derived  in a general  parametrization in order to exhibit  the diffeomorphism of 5d action.  For phenomenology study,  when one  sets  $F'- A' G =0 $ and $G = 2F$,   Eq.({\ref{3F}-\ref{2hF}}) are consistent with  the vertices provided in~{\cite{deGiorgi:2020qlg}}.  But  for 3-radion vertices,  the paper~{\cite{deGiorgi:2020qlg}}  omitted the last term in Eq.({\ref{3F}), which is crucial for preserving the diffeomorphism.

\section{Invariance of graviton vertices}~\label{Appendix3} 
For the cubic couplings with two gravitons and one radion,  we can demonstrate  the  on-shell  invariance  after the radion stabilization.   According to  the derivation  in Section~\ref{invariance},   the second term in  Eq.(\ref{2hF}), in combination with a quadratic graviton term,  maintains its invariance under  the $\delta_\epsilon$ diffeomorphism  in an off-shell manner. Here we  further show that  the  terms proportional to $F'-A'G $ and $A''$ in the second line of Eq.(\ref{2hF}),  exactly cancel the expansions from the GW scalar and  potential terms in Eq.(\ref{Act}). 
\beq
&&  \mathcal{L}_{h^2}^{(3)}  =   \frac{1}{2 \kappa^2} \int dy e^{-4A}  \Big\{  3 \left[  \left( h'_{\mu \nu} h^{\mu \nu} -\frac{1}{2} h' h \right) \left( F' - A' G \right) + 2  \left(h_{\mu \nu} h^{\mu \nu} -\frac{1}{2} h^2 \right) F A'' \right]    \nonumber \\ && ~ +  \frac{\kappa^2}{2}  \left(h_{\mu \nu} h^{\mu \nu} -\frac{1}{2} h^2 \right) \Big[ \phi'_0 \varphi'  - 4F \phi'^2_0 + \frac{\partial V}{\partial \phi_0} \varphi   - \sum_i \left( 4 F \lambda_i  - \frac{\partial \lambda_i}{\partial \phi_0} \varphi  \right) \delta(y - y_i) \Big]    \Big\}  \, \label{L2h}
\eeq
Using the gauge $F' - A' G = \frac{\kappa^2}{3} \phi'_0 \varphi$, we can convert the first term in  Eq.(\ref{L2h}) to be:
\beq
&& \frac{1}{2 } \int dy e^{-4A}   \left( h'_{\mu \nu} h^{\mu \nu} -\frac{1}{2} h' h \right) \phi'_0 \varphi \nonumber\\
&=& -\frac{1}{4} \int dy e^{-4A}  \left( h_{\mu \nu} h^{\mu \nu} -\frac{1}{2} h^2  \right)  \left( \phi''_0 \varphi + \phi'_0 \varphi'  - 4 A' \phi'_0 \varphi \right) \nonumber \\
&=&  -\frac{1}{4} \int dy e^{-4A}  \left( h_{\mu \nu} h^{\mu \nu} -\frac{1}{2} h^2  \right)  \left(  \frac{\partial V}{\partial \phi_0} \varphi +  \phi'_0 \varphi' + \sum_{i} \frac{\partial \lambda_i}{\partial \phi_0} \varphi \, \delta(y-y_i)  \right) \, \label{L0}
\eeq
where  we  perform the partial integration, and use Eq.(\ref{BG0}) to simplify the expression. Then substituting Eq.(\ref{L0}) into Eq.(\ref{L2h}) gives:
\beq
\mathcal{L}_{h^2}^{(3)}  &=&   \int dy e^{-4A}  \left(h_{\mu \nu} h^{\mu \nu} -\frac{1}{2} h^2 \right) \Big[ \frac{3}{\kappa^2} F A''  -F \phi'^2_0  - \sum_i  F \lambda_i  \, \delta(y - y_i) \Big]  
\eeq
that is identically zero after applying the BG equation (\ref{BG1}). Hence our result  indicates  that  as long as Eq.(\ref{gauge})   and  equations of  the background metric  are satisfied, the on-shell  diffeomorphism is still conserved for the effective Lagrangian with two gravitons.